\documentclass{aastex}
\usepackage[]{emulateapj5}
\usepackage[]{epsfig}
\tightenlines

\font\boldsym=cmmib10

\def    \bomega {{\hbox{\boldsym\char'041}}}	%bold \omega
\def	\be	{\begin{equation}}
\def	\ee	{\end{equation}}
\def    \ba     {\begin{eqnarray}}
\def    \ea     {\end{eqnarray}}
\def	\Angstrom	{\,{\rm \AA}}		% Angstrom
\def	\G	{{\rm G}}

\def	\K	{{\rm K}}

\def	\bH	{{\bf H}}

\def	\bJ	{{\bf J}}

\def	\ba	{{\bf a}}

\def	\cm	{\,{\rm cm}}

\def	\g	{\,{\rm g}}

\def	\s	{\,{\rm s}}
\def    \simlt  {\lower.5ex\hbox{$\; \buildrel < \over \sim \;$}}
\def    \simgt  {\lower.5ex\hbox{$\; \buildrel > \over \sim \;$}}
\def	\gtsim	{\simgt}
\def	\ltsim	{\simlt}

\def\pmb#1{\setbox0=\hbox{#1}%
\kern-.025em\copy0\kern-\wd0	
\kern-.05em\copy0\kern-\wd0
\kern-.025em\raise.0433em\box0}

\lefthead{Lazarian \& Draine}
\righthead{Resonance Relaxation}

\begin{document}

\title{Resonance Paramagnetic Relaxation and Alignment of Small Grains}

\author{A. Lazarian \& B. T. Draine}
\affil{Princeton University Observatory, Peyton Hall, Princeton,
NJ 08544}

\begin{abstract}

We show that the energy-level splitting arising from grain rotation ensures 
that paramagnetic dissipation acts at its maximum rate,
i.e.,
the conditions for paramagnetic resonance are automatically fulfilled.
We refer to this process as ``resonance relaxation''.
The differences between the predictions of classical
Davis-Greenstein relaxation
and resonance relaxation are
most pronounced for grains rotating faster than 1~GHz, i.e.,
in the domain where classical paramagnetic relaxation 
is suppressed.
This mechanism can partially align even very small grains,
resulting in linearly polarized 
microwave emission which could interfere 
with efforts to measure the polarization of the cosmic
microwave background.

\end{abstract}


\section{Introduction \label{sec:intro}}

Experiments to study the cosmic background radiation 
have stimulated renewed interest in diffuse galactic emission.
Recent 
maps of the microwave sky brightness have revealed a component of the 10-100
GHz microwave continuum which is correlated with 100 $\mu$m thermal 
emission from interstellar dust 
(see review by Draine \& Lazarian 1999a). 
Draine \& Lazarian (1998a,b, henceforth DL98a,b) attributed
this emission to electric dipole radiation
from small ($<10^{-7}$~cm) rapidly rotating grains.
Recent observations by de Oliveira-Costa et al.\ (1999) support this
interpretation.
The question now is whether these small grains are aligned
and their emission polarized.

One process that might produce alignment of the ultrasmall
grains is the
paramagnetic dissipation mechanism suggested 
%A23
%half a century ago
by Davis and Greenstein (1951) to explain the
polarization of starlight. The Davis-Greenstein mechanism is
straightforward: 
the component of interstellar magnetic field
perpendicular to the grain angular velocity varies in grain coordinates,
resulting in time-dependent magnetization, 
energy dissipation, and a torque acting on the grain.
As a result grains  tend to rotate
with angular momenta parallel to the 
interstellar magnetic field.

Although recent research 
(Draine \& Weingartner 1996, 1997, Lazarian \& Draine 1999a,b) 
suggests that paramagnetic alignment may not be the dominant 
alignment mechanism for 
$a\gtsim 10^{-5}\cm$ grains,
it may be effective for small 
($a\ltsim 5\times10^{-6}$~cm) grains.
%ALR (we repeat this later in section 2)
%For rotational velocity $\omega$,
%the paramagnetic alignment timescale $\tau\propto a^2/K$, where
%$K(\omega)\equiv {\rm Im}(\chi)/\omega$ for
%magnetic susceptibility $\chi(\omega)$
%For $\omega \leq 10^8$~s$^{-1}$, normal materials at $T\approx 20$~K
%have $K\approx 10^{-13}$~s. At higher frequencies, however, $K(\omega)$
%decreases rapidly (Draine \& Lazarian 1999b, henceforth DL99b). 
%Very small ($a\leq 10^{-6}$~cm)
%grains may rotate at very high frequencies (DL98b)
%thus calling into 
%question the efficacy of alignment by paramagnetic dissipation.
%

In the present paper we claim that the traditional picture
of paramagnetic relaxation is 
incomplete, as it
disregards the splitting of energy levels that arises within a rotating
body. 
Unpaired electrons having spin parallel and antiparallel
to the grain angular velocity have different energies resulting in the
Barnett effect (Landau \& Lifshitz 1960) --
the spontaneous magnetization of a paramagnetic body
rotating in field-free space. 
Therefore the implicit assumption in Davis \& Greenstein (1951) 
-- that the magnetization within a {\it rotating grain} in a {\it static} 
magnetic field is equivalent to the magnetization within a 
{\it stationary grain} in a {\it rotating} magnetic field --
is clearly not exact.  

In what follows we show that a very important effect due to rotation has
thus far been overlooked.
%btd 00.04.19 -----------------------
%This effect -- which we term ``resonance relaxation'' --
%leads to energy dissipation 
This effect, which we term ``resonance relaxation'',
leads to energy dissipation 
-- and grain alignment -- 
%------------------------------------
which is much more rapid than the
classical Davis-Greenstein estimate when the grain rotates very rapidly.

\section{Davis-Greenstein Theory}

%Even in the absence of paramagnetic ions, cosmic irradiation is
%likely to produce a high concentration of free radicals 
%(Greenberg 1982; Kaiser et al.\ 1997). 
%Carbon rings in coals exhibit
%paramagnetism (Overhauser 1953; Al'tshuler \& Kozyrev 1964)
%as do small hydrocarbon molecules (Atherton 1973).
%Thus we expect the population of small grains 
%to be paramagnetic, although the actual
%concentration of paramagnetic species is uncertain.

%A23
%Davis-Greenstein theory assumes that the dissipation caused by a stationary
%magnetic field in a rotating paramagnetic grain is the same as that arising
%from an applied
%magnetic field rotating in a stationary grain. 
Paramagnetic dissipation in a {\it stationary grain} 
depends upon the imaginary 
part of the magnetic susceptibility $\chi^{\prime\prime}$, 
which characterizes the
phase delay between the grain magnetization and the rotating
magnetic field. Due to this delay 
%btd 00.04.19 ------------------
%	we have just discussed a stationary grain.  It is confusing to
%	now say that ``the grain experiences a decelerating torque''.
%the grain 
a grain rotating in a static magnetic field
%-------------------------------
experiences a decelerating torque:
the energy dissipated in the grain comes from 
rotational kinetic energy.
The Davis-Greenstein alignment time scale is
\be
\tau_{\rm DG} \approx 3\times10^2{\rm \,yr}\left(\frac{a}{10^{-7}\cm}\right)^2
\left[\frac{10^{-13}\s}{K(\omega)}\right]
\left(\frac{5\mu{\rm G}}{B_0}\right)^2
\ee
where
$K(\omega)\equiv\chi^{\prime\prime}(\omega)/\omega$ 
(see Davis \& Greenstein 1951). 
%btd 00.04.19 ------
% The precise behavior of $K(\omega)$ is 
%uncertain (see Draine 1996, DL99b).
%-------------------
Following DL99b we estimate
\be
K(\omega)\approx \frac{\chi_0\tau_2}{[1+(\omega \tau_2)^2]^2} 
\approx 10^{-13}{\rm\, s}
\frac{(20\K/T_d)}{[1+(\omega \tau_2)^2]^2} 
%-------------------------------------
~~~.
\ee

%ALR (response to the referee)
Very small grains are expected to be paramagnetic both due to
presence of free radicals, paramagnetic carbon rings (see Altshuler
\& Kozyrev 1964) and captured ions.
The spin-spin coupling time
\be
\tau_2\approx \frac{\hbar}{3.8 n_p g\mu_{\rm B}^2} \approx 
2\times10^{-9} \left( \frac{10^{21}\cm^{-3}}{n_p}\right)\s
\label{eq:tau_2}
\ee
(see 
%btd 00.04.19 ---
%	eliminate Draine 1996 from refs; refer to DL98b
% Draine 1996) 
DL98b)
%----------------
where $\mu_{\rm B}$ is the Bohr magneton, and
$n_p\approx 10^{21}$~cm$^{-3}$ is
the concentration of unpaired electrons, greater than in
coals (Tsvetkov et al 1993), but less than the concentration of 
free radicals envisaged by
Greenberg (1982). 
Eq.~(\ref{eq:tau_2}) then predicts 
a cut-off frequency $\nu_{\rm cut}=(2\pi \tau_2)^{-1}
\approx 0.1$~GHz.
In the extreme case where
$\sim$10\% of the atoms are paramagnetic
one can get $\nu_{\rm cut}$ as large as
1~GHz but hardly any higher.

%ALR
%The classical discussion of
%paramagnetic relaxation therefore fails to provide
%appreciable alignment of grains rotating at  $\nu>5$~GHz (
%%A23
%%Spitzer 1978,
%%Lazarian 1995, 
%%Draine 1996).
%see Draine 1996).
%We will now reexamine paramagnetic dissipation in a rotating grain and
%show that a very important effect due to rotation has thus far
%been overlooked.

%ALR
\section{Barnett effect and Bloch equations}
%\section{Barnett effect and energy level splitting}

The Barnett effect states that a body 
rotating with velocity $\bomega$ develops a magnetization
\be
{\bf M}=-\chi_0 \frac{\hbar \bomega}{g \mu_B} \equiv
\chi_0 \bH_{\rm BE}~~~.
\label{barnett}
\ee
where $\chi_0$ is the static susceptibility,
and
%ALR 
$H_{\rm BE}$ is the
%$H_{\rm BE}=\hbar\omega/g\mu_{\rm B}$ is the
%
``Barnett-equivalent'' field.
The essence of the Barnett effect is easily understood:
a rotating body can decrease its energy, while
keeping its angular momentum constant, if some of the angular momentum is
taken up by its  unpaired spins. By flipping one spin of angular
momentum $\hbar/2$, the system can reduce its rotational kinetic energy
by $\hbar \omega$.
%ALR
%The fraction of the spins which become aligned is determined by
%minimizing the free energy. 

Although
the Barnett effect has been long known in physics 
(see Landau \& Lifshitz 1960),
its importance in the context of interstellar grains was only appreciated
%btd 00.02.23
%rather 
%----------
recently (Dolginov \& Mytrophanov 1975; Purcell 1979;
Lazarian \& Roberge 1997; Lazarian \& Draine 1997, 1999a).
In the present paper we discuss a hitherto-unrecognized 
aspect of the Barnett effect, namely its influence on paramagnetic
dissipation in a rapidly rotating grain.

%A23
%Resonance relaxation is a physical effect that can be rigorously
%described without 
%%btd 00.01.21
%%any 
%%-------------
%reference to the effect of paramagnetic resonance.
%However, to make our presentation easier we will 
%%btd 00.01.21
%%frequently 
%%------------
%refer the reader to the extensive literature  on 
%electron paramagnetic resonance,
%since the phenomena that we will describe have analogs
%described there. 

%btd 00.04.19 shorten ---
%In terms of quantum mechanics the presence of rotation 
Rotation
%-------------------------
removes the spin degeneracy of the electron energy levels. 
The energy difference between 
%btd 00.04.19 reword -----
%spin orientation parallel and anti-parallel to the total angular velocity
electron spin parallel or antiparallel to $\bomega$
%--------------------------
provides a level splitting corresponding to $\hbar \omega = 
g\mu_{\rm B}H_{\rm BE}$. 
%A23
%Finite coupling of electrons with the lattice
%enables the electrons with higher energy to jump to a lower
%level and the paramagnetic medium becomes magnetized. 
Insofar as the energy
levels and magnetization are concerned, rotation of the grain is analogous to 
%btd 00.02.23
%the 
%------------
application
of the ``Barnett equivalent'' field. 
%btd 00.04.19 ---------------
%	The reader could be confused if they thought we were discussing
%	an applied field parallel to the rotation.  At the cost of adding
%	words, I think that we have to make the geometry clear.
%	Start a new paragraph.
%In grain coordinates, a (weak) static magnetic field 
%%btd 00.02.23
%%now 
%%--------------
%appears like an applied field rotating 
%with frequency
%$\omega$. 

Now consider a (weak) static magnetic field $\bH_1 $ 
at an angle $\theta$ to $\bomega$.
In grain coordinates, this appears like a static field $H_1\cos\theta$
plus a field $H_1\sin\theta$ rotating with frequency $\omega$.
%-------------------
This rotating field 
can be resonantly absorbed, since the energy level 
splitting is exactly $\hbar \omega$.
%ALR
%This energy absorption, of course, is proportional to 
%$\chi^{\prime\prime}(\omega)$.
%In the classical Davis-Greenstein analysis this magnetic susceptibility
%$\chi(\omega)$ is taken to be that of a sample at rest in zero magnetic
%field: $\chi\equiv \chi(H_0=0, \omega)$. Here we point out that one should
%instead use $\chi=\chi_{\bot}(H_0=H_{\rm BE}, \omega)$, where 
%$\chi_{\bot}(H_0,\omega)$ describes the response of nonrotating material 
%to a weak field rotating at
%frequency $\omega$ perpendicular to a static magnetic field $H_0$.
%
%\section{Bloch equations}

The Bloch equations (Bloch 1946) 
are useful for describing
both resonant and nonresonant absorption (see Pake 1962). 
These phenomenological
equations reflect the tendency of the magnetization ${\bf M}$
to precess and
to tend exponentially towards its thermal equilibrium value 
${\bf M}_0$. In a stationary grain ${\bf M}_0=\chi_0 
{\bf H}$, where $\chi_0$ is the static 
paramagnetic susceptibility and  $\bf H$ is the
external field. In the case of a rotating grain the magnetization
arises due to the Barnett effect and 
%btd 00.01.21
is
%--------------
therefore given by eq.~(\ref{barnett}).

In what follows we assume that the grain magnetization is directed
along the $z$-axis and is dominated by
the Barnett 
%btd 00.02.23
%effect. Indeed, the
%``Barnett equivalent'' magnetic field for a grain rotating at
%$20$~GHz is $3.6\times 10^4$~G, much greater than 
%the internal magnetic field
%in a paramagnetic grain.
effect (the magnetic field for a grain rotating at
$20$~GHz is 7.2kG, much greater than 
the internal field
in a paramagnetic grain).
%-----------------------------
%A23
%Therefore
%one may ignore variations of the magnetization 
%direction introduced both by internal and interstellar magnetic fields.
%
%It is assumed that there exists an axis of strong magnetization,
%taken to be the $z$-axis. 
Changes in ${\bf M}$ along this axis involve changes in the 
energy of the spin 
%btd 00.02.23
%system and thus 
system, thus
%----------
require spin-lattice interactions, 
and therefore occur on the spin-lattice relaxation time scale $\tau_1$ 
(Atherton 1973). 
Changes in ${\bf M}$ in the perpendicular
direction only slightly perturb its direction but not its magnitude.
The  interactions within the electron spin
system are sufficient to deflect the direction of magnetization and
these perturbations relax on the spin-spin relaxation time $\tau_2$.

%A23
%we choose the $z$-axis along ${\bomega}$,and 
Using $\bot$ for the $x$ and $y$ components, 
the Bloch equations in the presence of the interstellar
magnetic field ${\bf H}_1$ are (see Morrish 1980)
\be
\left(\frac{d}{dt}\right){\bf M}_{\bot}+[\bomega\times {\bf M}]_{\bot}=
\gamma g\left[{\bf H}_{1}\times {\bf M}\right]_{\bot}-\frac{{\bf M}_{\bot}}{\tau_2}~~~,
\label{bot}
\ee
\be
\left(\frac{d}{dt}\right)M_z
 =\gamma g\left[{\bf H}_{1}\times {\bf M}\right]_z+\frac{M_0-M_z}{\tau_1}~~~,
\label{z}
\ee
where  $d{\bf M}_{\bot}/ dt$ represents the motion of ${\bf M}_{\bot}$
in 
body coordinates and $\gamma\equiv e/2 m_e c=8.8\times 10^6$~s$^{-1}$G$^{-1}$.
%Eqs.~(\ref{bot}) and (\ref{z}) use the well known formulae to
%relate the dynamics of $\bf M$ in the 
%inertial (primed) and rotating systems of reference:
%\be
%\left(\frac{d}{dt'}\right){\bf M}=\left(\frac{d}{dt}\right){\bf M}+
%\bomega \times {\bf M}~~~.
%\ee

%btd 00.02.25 delete following sentence and add ``stationary'' to next
%Interstellar grains rotate in a stationary interstellar
%magnetic field H$_1$.
%--------------------------------------
%btd 00.02.23
%We will discuss the magnetic response in the system of reference 
%$({\bf i},{\bf j},{\bf k})$ rotating with 
%the grain at angular velocity $\bomega$. In this system of reference
%rotating with the grain 
%the component of {\it interstellar} magnetic field perpendicular to $\bomega$
%is
Consider a reference frame
$({\bf i},{\bf j},{\bf k})$ rotating with 
the grain at angular velocity $\bomega$. In this frame
the 
%btd 00.02.25 add stationary here
stationary
%---------
{\it interstellar} magnetic field 
is
%-----------------
\be
{\bf H}_1=[{\hat{\bf i}}H_1\cos(\omega t)
%btd 00.02.25
% 		correct sign error
%+ 
-
%---------------------------
{\hat{\bf j}}H_1\sin(\omega t)]\sin\theta
+{\hat{\bf k}} H_1\cos\theta~~~.
\ee
%btd 00.04.19 --------------------
%	change phi to theta, and delete definition of angle
%	since we have defined it above
%where $\phi$ is the angle between $\bomega$ and ${\bf H}_1$.
%As the magnetization in the $z$-direction $\chi_0 H_1 \cos\phi$
As the magnetization in the $z$-direction $\chi_0 H_1 \cos\theta$
%----------------------------------
is much smaller than 
%btd 00.04.19 delete to save space-----
%the Barnett magnetization 
%--------------------------------------
$\chi_0 H_{\rm BE}$,
it is disregarded in our treatment. 

%ALR
%
For  $M_z$ the stationary solution is $M_z=-\chi_0 H_{\rm BE}/f_{\rm st}$
where $f_{\rm st}=1+\gamma^2 g^2\tau_1 \tau_2 H^2_1\sin^2\theta$, 
%The stationary solution can be obtained for $M_z$:
%\be
%M_z=-\frac{\chi_0 H_{\rm BE}}{1+\gamma^2 g^2\tau_1 \tau_2 H^2_1\sin^2\phi}~~~,
%\ee
while $M_x$ and $M_y$ oscillate with a $\pi/2$ lag with respect  to the
interstellar magnetic field. For instance,
%ALR
$M_x=\chi_0 \omega \tau_2 H_1
\sin(\omega t)/f_{\rm st}$.
%\be
%M_x=\frac{\chi_0 \omega \tau_2 H_1
%\sin(\omega t)}{1+\gamma^2 g^2\tau_1
% \tau_2 H_1^2\sin^2\phi}~~~.
%\ee
Therefore
%btd 00.04.19 --------------
%	Alex -- I think that it may be clearer if we write out the denominator
%	in eq. 8, even though this costs a bit of space
\be
%\chi''=\chi_0\omega \tau_2/f_{\rm st}
\chi^{\prime\prime}=\chi_0 
\frac {\omega \tau_2}
{1+\gamma^2 g_s^2 \tau_1 \tau_2H_1^2\sin^2\theta}~~~,
\label{eq:chi''}
\ee
which coincides with the expression for 
%btd 00.04.19 abbreviate----------------------
%the imaginary part of the susceptibility 
$\chi^{\prime\prime}$
%---------------------------------------------
for electron paramagnetic resonance
%btd 00.04.19 add words for clarity ------
in a stationary sample 
%-----------------------------------------
when the frequency of the oscillating field $H_1$ is equal to
%btd 00.01.21
%with 
%-------------
the resonance frequency. In our problem the only relevant
frequency is the frequency of grain rotation. Therefore it is not
accidental that the paramagnetic relaxation is ``resonant'' when the grain
rotates in the external magnetic field. The term $\gamma^2 g \tau_1
\tau_2 H_1^2\sin^2\theta$ in the denominator allows for 
``saturation'' when the energy dissipated in the spin system raises its
temperature due to slow spin-lattice coupling;
below we show that this term can be important
for very small grains, for which $\tau_1$ can be large.

%A23
%Our treatment above fails when the 
%magnetic field arising from interactions with
%neighboring spins is comparable with the ``Barnett equivalent'' magnetic
%field, just as the analogous ``strong field'' approximation in paramagnetic
%resonance experiments is incorrect when the local magnetic field deviates
%substantially from the direction of the external magnetic field. One may
%consult Pake (1962) how to obtain a treatment of the paramagnetic
%relaxation when spin precession is dominated by internal magnetic fields.
%A similar treatment applied to the rotating grain may be shown to give
%two separate contributions to $\chi^{\prime\prime}$, one of which scales as 
%$(\omega\tau_2)^{-3}$ for $\omega\tau_2\gg 1$ while the other coincides with 
%Eq.~(\ref{eq:chi''}).
%The former may be identified with the ordinary ``Davis-Greenstein
%relaxation'' as it is present 
%%btd 00.01.21
%%when grain is stationary while magnetic field is rotating, 
%for the magnetic field rotating in a stationary grain,
%%--------------------------------
%while the latter constitutes the resonance relaxation term.
%Both terms are of the same order for $\omega\tau_2< 1$, but it is the
%resonance term  that survives for  $\omega\tau_2\gg 1$. Being interested
%in relaxation within rapidly rotating grains here we discuss the resonance
%term only.

An important difference between paramagnetic resonance in an external
magnetic field and the resonance relaxation 
%btd 00.02.23
%as we discuss it 
discussed
%--------------
here is
that the ``Barnett equivalent'' magnetic field  
%btd 00.02.23
%acting on species with different magnetic moments is different. 
is different for species with different magnetic moments.
%---------------
Therefore, unlike paramagnetic
resonance, resonance relaxation happens simultaneously to species with
completely different magnetic moments and $g$-factors. 
For example, the conditions for electron spin resonance 
and nuclear magnetic resonance are satisfied simultaneously when a grain 
rotates in a static 
%btd 00.02.23
weak
%-------
magnetic field.

\section{Spin-Lattice Relaxation}

For a spin to flip in a rotating grain, it is necessary for the
total energy in lattice vibrations to change by $\hbar \omega$.
Because the density of states is finite, it may not be possible
for the lattice vibrations to absorb an energy 
$\hbar \omega$. 

To estimate 
%btd 00.01.17
the lowest vibrational frequency
%-------------
$\omega_{\min}$ we note that the lowest frequency bending
mode of 
the coronene molecule (C$_{24}$H$_{12}$) is 
%btd 00.04.19 delete words----
%estimated to be at 
%-----------------------------
$\omega_{\min}=1.9\times10^{13}\s^{-1}$ 
%btd 00.02.23
%($\lambda^{-1}= 102\cm^{-1}$) (Cyvin 1982; Cyvin etal 1984).
(Cyvin 1982).
Coronene
has an effective radius 
$a=4.3\Angstrom$ for an assumed density $\rho=1.5\g\cm^{-3}$; if 
$\omega_{\min}\propto a^{-1}$, then 
$\omega_{\min}=8.3\times10^{12} a_{-7}^{-1} \s^{-1}$,
%ALR
%\be
%\omega_{\min}=8.3\times10^{12} a_{-7}^{-1} \s^{-1} ~~~,
%\ee
large compared to $kT/\hbar$.
%btd 00.04.19 reword ----------
%Thus one cannot use 
%the measured spin-lattice relaxation times that are provided
%for macroscopic bodies while dealing with microscopic grains.
Thus for ultrasmall grains one cannot use 
spin-lattice relaxation times $\tau_1$ measured for macroscopic samples.
%-------------------------------

We obtain an upper estimate for $\tau_1$ by 
appealing to the Raman scattering of phonons (see
Waller 1932, Pake 1962): annihilation of a vibrational quantum
$\hbar \omega'$ and creation of a quantum $\hbar (\omega - \omega')$.
Elastic lattice vibrations  of 
all frequencies participate and therefore the relaxation is
present for small grains. 
After integrating over various modes of
vibrations one gets the probability of the spin-lattice transition via
Raman scattering (Al'tshuler \& Kozyrev 1964)
\be
\tau_1^{-1}\approx K_2 \left(kT/\hbar\right)^{m+1}
\tilde{J}_m/\rho c_s^{10}~~~,
\label{T_K}
\ee
where $\rho$ is the 
%btd 00.04.19 shorten --- 
%density of the material, 
density,
%------------------------
$c_s$ the sound speed,
$K_2$ is a function that depends on the density of states, and
$\tilde{J}_m$ ($m=$6 or 8) is an integral over the body's phonon frequencies:
$$
\tilde{J}_m = \int_{T_l/T}^{\theta_D/T} \frac{x^m e^x}{(e^x-1)^2} dx ~~,
$$ 
where $\theta_D$ is the Debye temperature.
The conventional treatment assumes the body to be infinite
with the integration extending down to $T_l = 0$.  In this case,
and for $T\ll\theta_D/m$, we have $\tilde{J}_m\approx m!\zeta(m)$,
where $\zeta$ is the Riemann zeta function.
For a 
grain of size $a$,
\be
T_l = \frac{\hbar \omega_{\min}}{k} \approx \frac{63\K}{ a_{-7}} ~~.
\ee
For $T\ll T_l/m$ we have
$\tilde{J}_m\approx 
\left(T_l/T\right)^m
\exp(-T_l/T)$.
The ratio of the Raman spin-lattice relaxation in a small grain
at temperature $T \ll T_l/m$
to such relaxation in an infinite body at 
$77K \ltsim \theta_D/m$ is then
\be
\!\frac{\tau_{1}(T)}{
\tau_{1, \infty}(77\K)}\approx \left(\frac{77\K}{T}\right)^{m+1}
\!\!\left(\frac{T}{T_l}\right)^m \!\!\exp(T_l/T) m!\zeta(m) .
\label{eq:tau_1,grain}
\ee
Data in Al'tshuler \& Kozyrev (1964) suggests that ionic crystals
have a  spin-lattice relaxation time 
%btd 00.01.17
%of the order of $10^{-6}$~s at $77$~K. 
$\tau_{1,\infty}(77\K)\approx10^{-6}\s$.
If due to the 
Raman process, then we would estimate that a macroscopic sample
at $T=4\K$ would have $\tau_1\approx 28\s$ if $m=6$,
or $2\times 10^4\s$ if $m=8$.
>From eq. (\ref{eq:tau_1,grain}), a grain with $a=10^{-7}\cm$ at $T=4\K$ 
would have
$\tau_1 \approx 3.1\times10^5\s$ for $m=6$, or $2.7\times10^7\s$ for $m=8$.
Grains with $a < 10^{-7}\cm$ would have even larger values of $\tau_{1}$.

If we adopt $\tau_2$ from eq. (\ref{eq:tau_2}),
then
\be
g_s^2\gamma^2 \tau_2 \tau_1
H_1^2 \sin^2\theta
= 
%btd 00.02.23 factor of 5**2 correction
%6 \left(\frac{\tau_{1,grain}}{10^8\s}\right)
%btd 00.02.25 still numerically incorrect. fix:
%1.5 
8
%------
\left(\frac{\tau_{1,grain}}{10^6\s}\right)
%----------------
\left(\frac{H_1}{5\mu\G}\right)^2
%btd 00.04.19 changed phi to theta:
\frac{\sin^2\theta}{2/3}
\ee
so we see from (\ref{eq:chi''})
that saturation may be important for $a < 10^{-7}$
grains even in the $\sim 5\mu\G$ fields in diffuse interstellar gas.

How reliable 
%btd 00.01.17
%our estimate of $\tau_1$ above? 
is our above estimate for $\tau_1$?
%------------- 
Our calculations
were based on so-called Waller theory, which frequently
overestimates 
%btd 00.02.01 --------------
%the relaxation times 
$\tau_1$
%---------------------------
by a factor up to $10^8$ (Pake 1962).
If the dependence of the spin-lattice relaxation time on temperature
is different from that given by Eq.~(\ref{T_K}) our estimates of $\tau_1$
at $4K$ would  
be very different. 
%btd 00.02.01
%However, ultimately 
Ultimately
we require laboratory measurements of $\tau_1$ in
small particles of appropriate composition.
%-----------

\section{Grain alignment}

Paramagnetic alignment of grains with a given axis ratio
depends on two 
%btd 00.02.23
%parameters, namely,
parameters:
%-------------
the ratio  $T_{\rm d}/T_{\rm rot}$ of grain vibrational and
rotational temperatures, and 
%btd 00.02.23
%on the ratio of the paramagnetic relaxation time $t_{\rm r}$ to
the ratio of the alignment time $\tau_{\rm DG}$ to
the rotational damping time $t_{\rm d}$. 
%----------------------
The rotational damping time $t_{\rm d}$ depends
on various processes of damping and excitation 
(e.g. collisions with ions and neutrals, plasma drag, and
emission of photons) discussed in DL98b.
%btd 00.04.19 -----------------------------------------------------------------
%	I suggest deleting this, since we state that we compute \omega
%	following DL98b
%We take $\omega \approx 
%\left(45kT_{rot}/16\pi\xi\rho a^5\right)^{1/2}$,
%where $\rho\approx1.5\g\cm^{-3}$ is the grain density, 
%$a$ is the radius of an equal volume sphere,
%and
%$\xi$ is the ratio of the moment of inertia to that of
%an equal volume sphere.
%%btd 00.02.23 combine paragraphs
Assuming that the
paramagnetic torque only marginally 
%changes the rate of grain rotation 
reduces $\omega$,
we follow the analysis of DL98b to obtain 
%$\omega$, $T_{rot}$ and $\xi$ as functions of $a$.
$\omega$ as a function of $a$.
%-----------------------------------------------------------------------------

In our calculations we assume that 
a grain spends most of its time between thermal spikes with a vibrational
temperature $T\approx4\K$ (cf. Rouan et al. 1992), since
the time between photon absorptions is $\sim 10^9$~s,
%ALR
%while any vibrational transition with an electric dipole moment will decay
%much more rapidly.
while the grain cools much more rapidly. Note that photons usually
contribute marginally to the disorientation of the grain angular
momentum $\bf J$ (see Fig.\ 5 in  DL98b).

%ALR
Let $\theta$ be the angle between 
%btd 00.04.19 delete ------
%the grain 
$\bf J$ 
%--------------------------
and the
interstellar magnetic field $\bf B$.
Fig.~1 presents the measure of alignment 
$\sigma=\frac{3}{2}\langle\cos^2\theta-1/3\rangle$, 
for grains in the cold neutral medium.
For our estimate we used standard formulae for paramagnetic
alignment of angular momentum with $\bf B$
(Lazarian 1997, Roberge \& Lazarian 1999), which for weak alignment provide 
$\sigma\approx 2/15[1-(1+rt)/(1+r)]$, 
where $r\equiv 
%btd 00.02.23
%t_r/t_{\rm d}$
\tau_{\rm DG}/t_{\rm d}$
%--------
and $t\equiv T/T_{rot}$. 

The discontinuity 
at $\approx 6\times 10^{-8}$~cm
is due to the assumption that smaller grains are planar, and larger
grains are spherical.
%A23
%We also show the rms rotational frequency $\omega/2\pi$
%as a function of $a$. 
The degree of polarization of rotational electric dipole emission
$p\approx \sigma\cos^2\psi$, where $\psi$ is the angle between $\bf B$ and
the plane of the sky.
$p/\cos^2\psi$ as a function of frequency
is also shown in Fig.~1.
%
%------------------------------------------
The dipole rotational 
emission predicted in DL98a,b is sufficiently 
strong that polarization of a few percent
may interfere with efforts to measure the polarization of
the cosmic microwave background radiation.
%
%AL
It worth noting that the degree of microwave polarization is sensitive to
the magnetic field {\it intensity} 
%btd 00.02.01
%	I don't believe the following statement is correct.
%Thus ratio of the polarization
%degrees for physically similar regions may be used to find the
%relative value of the magnetic field component perpendicular to the
%line of sight.
%A23 
%(through $\tau_{\rm DG}$) as well as $\cos^2\psi$.
(through $\tau_{\rm DG}$).

\section{Discussion}

We have discussed a new gyromagnetic effect --
resonance relaxation --
which is closely related to normal paramagnetic resonance,
and arises naturally whenever a body rotates 
in a weak magnetic
field. 
%btd 00.02.23
%We showed that the 
The
%------------------
standard assumption of
the equivalence
of relaxation when the magnetic field rotates 
%A23
about a grain
or a grain rotates in 
a static magnetic
field is 
incorrect; the
difference is directly related to the 
spontaneous magnetization due to the Barnett
effect.

Although present for all grains, 
%btd 00.01.17
%the effect 
resonance relaxation 
%---------
is most prominent for the smallest ones. 
When grains rotate very 
%btd 00.02.23
%fast, 
rapidly,
%-------
%btd 00.04.19 shorten
%as is the case, for instance,
%for very small grains, the resonance relaxation effect
%ensures that the paramagnetic susceptibility 
as is the case for very small grains, the resonance relaxation effect
ensures that 
%------------------------
$\chi^{\prime\prime}$ 
does not plunge
as the rotation frequency increases. As a result, we conclude that
small grains (e.g. $a\leq 10^{-7}$~cm) should be paramagnetically 
aligned. 
The degree of their aligment 
depends
on the particular phase of the interstellar medium 
and on
the efficiency of spin-lattice relaxation. The latter factor
is unfortunately uncertain for very small grains for which the existing
laboratory data is not applicable.

%btd 00.04.19 I substantially edited the following paragraph...
If the ultrasmall grains are partially aligned, the implications
are as follows:
(1) The microwave radiation described in DL98ab
will be polarized -- by a few \% -- and could have dramatic consequences for
experiments -- such as MAP or PLANCK --
designed to measure polarization of the cosmic microwave 
background.
(2) If the grain body axes are aligned with $\bJ$, then absorption by
these small grains will contribute to starlight polarization in
the ultraviolet, 
and (3) the infrared emission following 
absorption of starlight photons by these small grains will also
be polarized.  
However, the contribution to starlight polarization is expected to be
small due to only partial alignment of
the grain body axes with $\bJ$.
The infrared emission will be even less polarized, due to
disorientation of the grain axes (Lazarian \& Roberge 1997) 
during the thermal spike following a photon absorption, i.e.,
while the infrared emission is taking place.

\acknowledgements
This work was supported in part by NASA grant NAG5 7030 and in part by
NSF grant AST-9619429.
%btd 00.02.23
%Anew
We thank ...
% ``unveil'' these after paper is accepted:
%Anew
J. Mathis and J. Weingartner 
% Alex: who else?  Suggest we keep them in alphabetical order.
for valuable discussions.
%------------------------------
%A23
%A.L. is grateful to Peter Martin, Frank Shu,
%Paul Steinhardt and 
%Paul Shapiro for valuable discussions
%of the results.

%btd 00.02.23
%	Alex -- should not try to specify epsfysize separately from
%	epsfxsize, since this will end up distorting the symbols in 
%	the figure.  Have set epsfxsize here to 0.45\columnwidth as
%	this still keeps the figure on page 4.
%Anew
% we should return back to smaller y-dimension, I believe.
%--------------
%ALR
\begin{figure}[h]
{\centering \leavevmode
\epsfxsize=.45\columnwidth 
\epsfysize=.35\columnwidth
\epsfbox{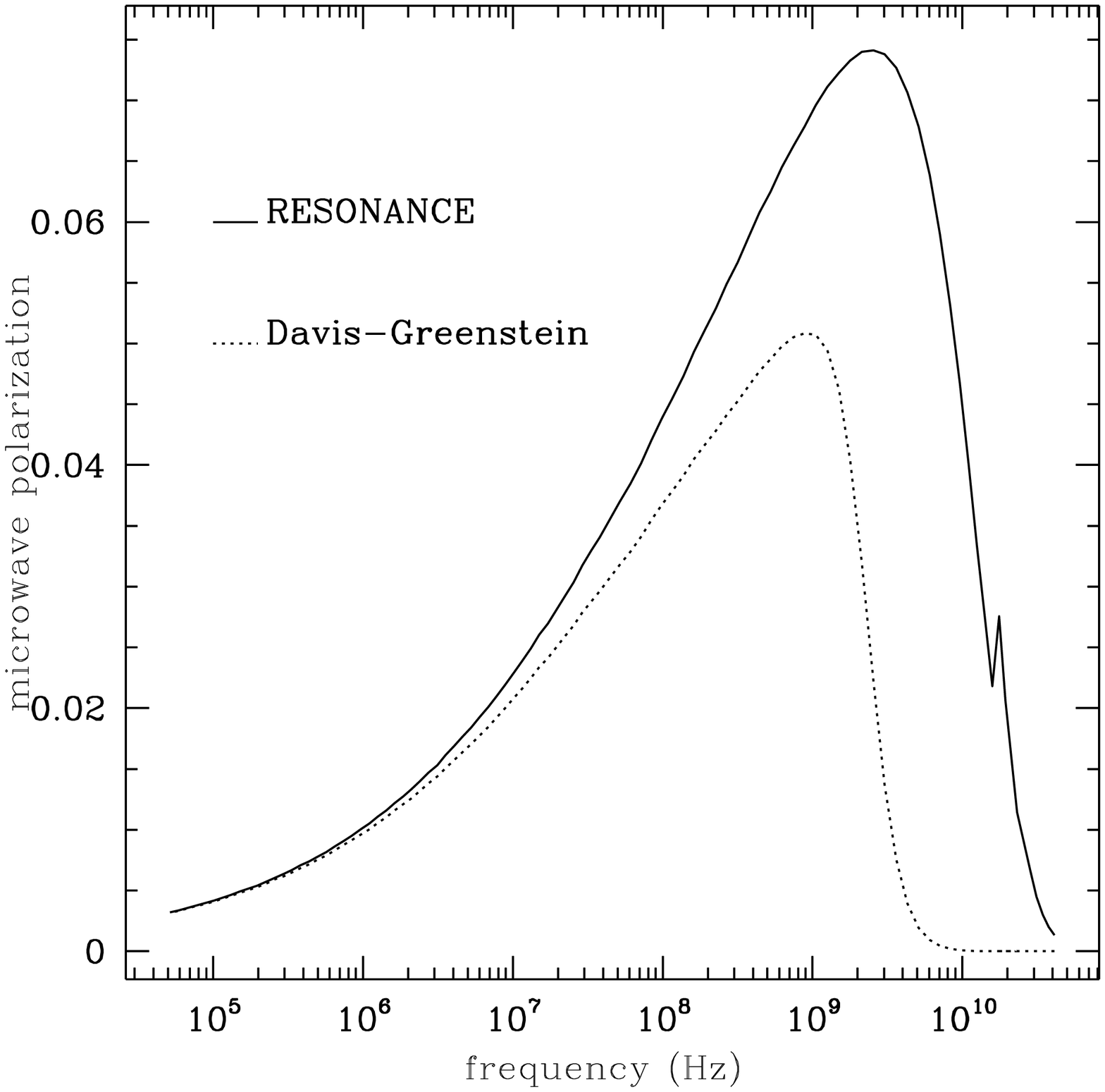} \hfil
\epsfxsize=.45\columnwidth
\epsfysize=.35\columnwidth 
\epsfbox{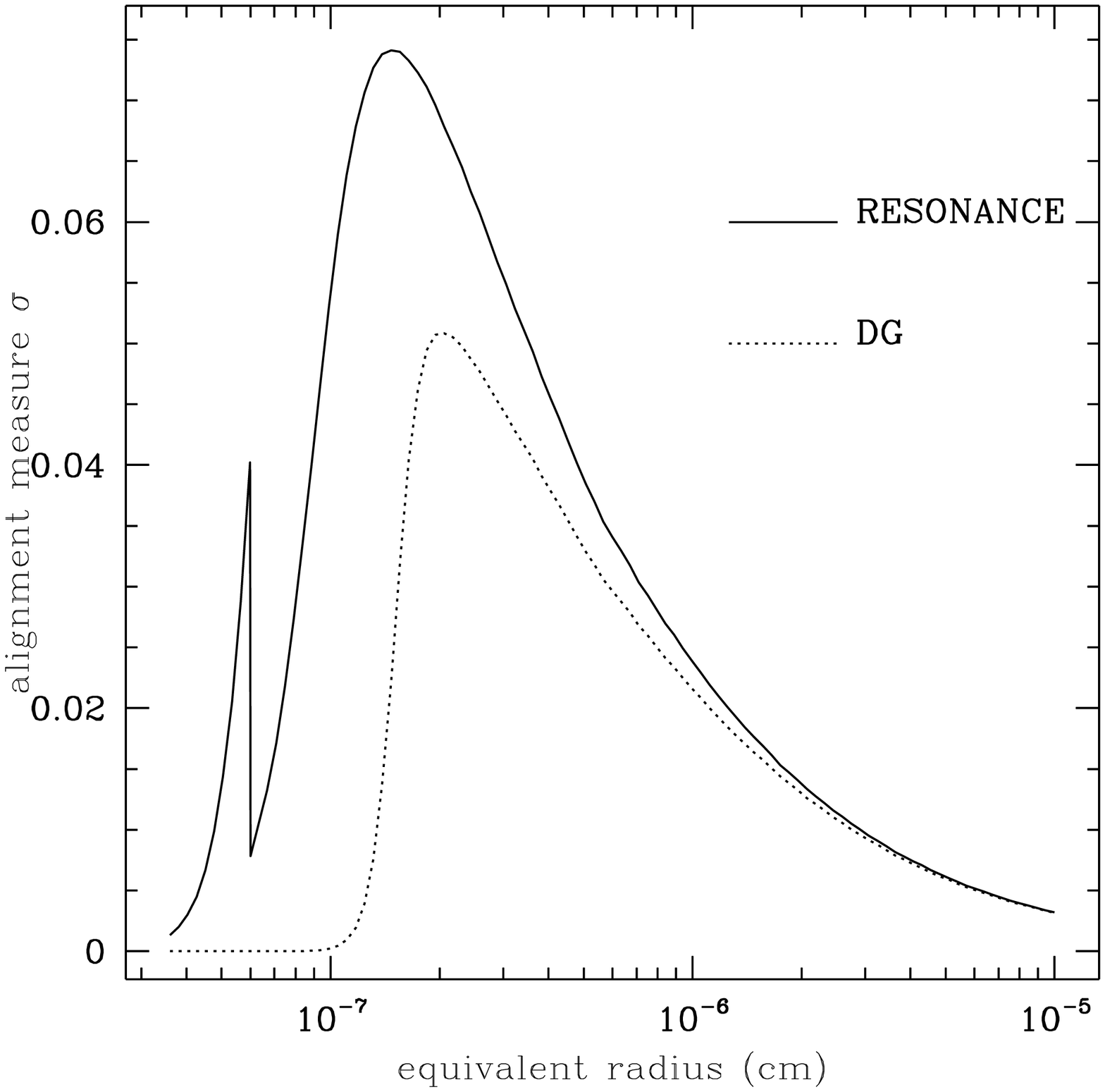} \hfil
}
\caption{Measure of grain alignment for both
	resonance relaxation and Davis-Greenstein relaxation for grains in
	the cold interstellar medium as a function of frequency (left)
and size (right). 
%Anew
For resonance relaxation the saturation effects 
%btd 00.02.25
(see eq.\ \ref{eq:chi''}) are neglected, which means
that the upper curves correspond to the {\it maximal} values allowed by the 
paramagnetic mechanism.
The amplitude of discontinuity on the left is smaller than that on the
right as both spherical 
%btd 00.02.23
%grains and smaller flat 
and planar
%-----------------
grains contribute
to the microwave emissivity at 20-30~GHz.} 
\end{figure}

\end{document}